\documentclass[journal]{IEEEtran}
\usepackage{caption}
\usepackage{cite}
\usepackage{subcaption}
\usepackage[T1]{fontenc}
\usepackage{textcomp}
\usepackage{color,soul}
\usepackage{verbatim}
\usepackage{paralist}
\usepackage{amsthm}
\ifCLASSINFOpdf
  \usepackage[pdftex]{graphicx}
  \graphicspath{{../Figures/}}
\else
  \usepackage[dvips]{graphicx}
  \graphicspath{{../Figures/}}
\fi

\usepackage[cmex10]{amsmath}
\usepackage{amsfonts}
\usepackage{amssymb}

\usepackage{array}
\usepackage{url}
\usepackage{color,soul}

\newcommand{\norm}[1]{\left\lVert#1\right\rVert}
\newcommand{\RN}[1]{%
  \textup{\uppercase\expandafter{\romannumeral#1}}%
}

\newtheorem{Def}{Definition}
\newtheorem*{remark}{Remark}

\begin{document}

\bstctlcite{IEEEexample:BSTcontrol}
\newtheorem{definition}{\textbf{Assumption}}
\title{Data driven nonlinear identification of Li\textendash ion battery based on a frequency domain nonparametric analysis\\ \linespread{3}
 \footnotesize{\textcolor{blue}{ArXiV Preprint: The original article is published in IEEE Transactions on Control Systems Technology Volume 25, Issue 5, Sept. 2017  \\ DOI: 10.1109/TCST.2016.2616380}}} 
\author{Rishi~Relan\textsuperscript{*}, Yousef Firouz\textsuperscript{**}, Jean-Marc Timmermans\textsuperscript{**}, Johan~Schoukens\textsuperscript{*},~\IEEEmembership{Fellow,~IEEE} 

\thanks{The  authors\textsuperscript{*}  are  with  the  ELEC  Department and the authors\textsuperscript{**} are with MOBI research group of the Vrije  Universiteit  Brussel (VUB), Belgium respectively. Corresponding author can be contacted at Rishi.Relan@vub.ac.be. This work was supported in part by the IWT-SBO BATTLE 639,  Fund for Scientific Research (FWO-Vlaanderen), Methusalem grant, the Belgian Government through the Inter university Poles of Attraction (IAP VII) Program, and by the ERC advanced grant SNLSID, under contract 320378.}
}
\markboth{ IEEE TRANSACTIONS ON CONTROL SYSTEMS TECHNOLOGY}%
{Shell \MakeLowercase{\textit{et al.}}: Bare Demo of IEEEtran.cls for Journals}
\maketitle

\begin{abstract}
Lithium ion (Li-ion) batteries are attracting significant and growing interest because their high energy and high power density render them an excellent option for energy storage, particularly in hybrid and electric vehicles.  In this  paper, a data-driven polynomial nonlinear state-space model (PNLSS) is proposed for the operating points at the cusp of linear and nonlinear regime of the battery's electrical operation, based on the thorough nonparametric frequency domain characterization and quantification of the battery's behaviour in terms of its linear and nonlinear behaviour at different levels of the state-of-charge (SoC).  
\end{abstract}

\IEEEpeerreviewmaketitle

\IEEEpeerreviewmaketitle

\section{\textbf{Battery Modelling}}
\label{BattMod}
The pursuit for battery models with high accuracy and computational efficiency still remains a challenge. Generating a mathematical model of a Li-ion battey, e.g. needed by battery management system (BMS), that can describe the input current-to-output voltage dynamics of a battery is a challenging problem. A primary reason for this is that battery dynamics vary significantly with operating conditions. Depending on the final purpose of the developed model, one can divide battery models into the models that describe the short term behaviour e.g. state of charge, voltage response etc. and models that describe the long term behaviour of the cells, e.g. lifetime models, state of health etc. In the field of battery modelling many different battery models exist for both the short and long term behaviour of battery cells \cite{Cuma2015517}. These models can be broadly classified in the following categories: 
\begin{inparaenum}
\item Equivalent circuit models (ECM),
\item Electro-chemical  models,
\item Analytical and impedance based models,
\item Empirical and semi-empirical models. 
\end{inparaenum} 
ECMs are structurally simple and computationally efficient due to the use of lumped-parameter circuit elements, e.g. inductors, resistors and capacitors, to represent the battery impedance, and these models frequently incorporate empirical functions to describe the relationship between SoC and open circuit voltage (OCV). ECMs thus are widely used for impedance analysis\cite{Hang2013442}, SoC estimation \cite{HeHongwen2011} and charging control \cite{Hu2013449}. Performances of several commonly used ECMs is compared in \cite{hu2012comparative}. 
On the other hand electrochemical models \cite{Corno2015, Yiran2015, Stein2013} usually use coupled nonlinear partial differential equations to describe ion transport phenomena and electrochemical reactions to achieve high accuracy, but incurring heavy computation load. In general electrochemical models such as pseudo two dimensional models \cite{doyle1993modeling}, single particle models \cite{Bartlett2016}, and extended single particle models \cite{luo2013new} are more accurate than ECMs. In comparison ECMs are easier to implement, but have worse accuracy than electrochemical models \cite{rahimian2011comparison}, indicating that ECMs are unable to characterize battery impedance accurately due to their structural simplicity. Drawback of electrochemical models is that they require a large number of battery internal immeasurable parameters such as diffusion coefficients, concentration of species in electrolytes, electrode geometry and porosity, transfer coefficients, and reaction constant etc. to be estimated which leads to overfitting in a parametric identification. Hence, this approach is complex and difficult to use in practice. In analytical models, the major properties of batteries are modeled using few explicit equations to compute the battery states. However, such equations are not easy to solve. Peukert's law \cite{Jongkyung2011} is an example of such models. It captures the nonlinear relationship between battery lifetime and its rate of discharge, but without modeling the recovery effect. 

Empirical and semi-empirical models are good alternatives to the highly complex electro-chemical, electro-thermal or thermo-chemical models to describe the short-term behaviour of the battery eletrical response. In comparison to ECM and electro-chemical models, empirical methods do not explicitly rely on dedicated hardware/software and physics-based models of battery dynamics. If comparable and adequate training data are available under different operating conditions, data-driven methods are significantly more efficient than model-based methods in terms of computation, execution time, and memory requirements. In \cite{pattipati2011system} a  BMS framework is proposed that estimates the critical characteristics of the battery such as SoC, SOH, and Remaining useful life (RUL) using a semi data-driven approach. It used a combination of a modified Randles circuit model, support vector machines (SVMs), low-current Hybrid Pulse Power characterization (L-HPPC) test data, support vector regression, and a hidden Markov model (HMM). This method is quite good in nonlinearity mapping but it is very sensitive to the amount and quality of training data. A recursive least-squares  (Kalman filter) based system identification model for the online monitoring of batteries for electric vehicles (EVs) is developed in \cite{Juang2010}. The disadvantage is that the Kalman filter may be adversely affected by significant divergence problems when the battery model inaccurately reproduces the behavior of the battery.  A critical review of different methods for the monitoring of Li-ion batteries in HEVs can be found in \cite{Waag2014321}. A Fractional system identification is applied to lead acid battery state of charge estimation in \cite{sabatier2006fractional}. Fractional order systems accumulate the entire information of the system function in weighted form using a time varying initialization function which must be known as long as the system has been operated. Fractional dynamics require history of states or a sufficient number of points for the initialization function computation. Hence, it results in large memory requirements. Linear-parameter-varying (LPV) battery models for batteries used in HEV applications has been presented in \cite{hu2009technique}. Generally, LPV model suffer from serious disadvantages in terms of non-eliminatable  pitfalls of interpolation, selection of adequate linearization points, choice as well as estimation of a scheduling parameter, and the loss of a general representation of the nonlinear dynamics.
 Therefore, in order to develop a fast and an accurate dynamic model of the battery short term electrical response in different regions of its operating regime, the first and the foremost step must be to know, how the battery would behave in a particular scenario such as; at a particular setting of SoC, SoH and temperature i.e. when the battery would start operating in the nonlinear regime, when the time variations would become stronger etc. Hence, the main contributions in this paper is to show, how specially designed broadband excitation signals can be used to reveal useful information about the battery dynamics from the measured input current and output voltage signals. Once this information is available to the battery modeller, then it is discussed, how it can be exploited to model accurately different regimes of battery's operation. In the context of this paper, we use a complete black-box identification approach to identify the non-linear model of the battery dynamics utilizing the information about the nonlinear nature of the battery's dynamics gained from the nonparametric analysis. The advantage of using a black-box identification scheme is that it is a completely data-driven approach, and it does not require the user to have any pre-specified knowledge of the system. The structure and parameters of the models are learned from the data itself. 
 
 This paper is organized as follows: Section \ref{MultisineDef} introduces multisine excitation signals. Section \ref{NonChar} describes the reasoning behind, how the properties of specially designed multisine excitation signals can be utilized for the detection and quantification of nonlinearities in a dynamic system. Section \ref{NonMod} describes the nonlinear modelling approach using the PNLSS model proposed in this paper. The procedure for the identification of PNLSS model is described in Sectionc \ref{IdenPNLSS}. Section \ref{MeasExp} gives an introduction to the experimental set-up as well as the measurement methodology used for the acquisition of the signals under different SoC settings required for the analysis. Results of experimental investigations are presented in Section \ref{res}, and finally, the conclusions  are given in Section \ref{conc}.
\section{\textbf{Multisine as the perturbation signal}}
\label{MultisineDef}
 Before proceeding to model the battery dynamics, it is very important to characterize the battery electrical response under varying operational conditions in terms of the level as well as kind of non-linearities. It is possible to use a broadband \cite{al2013broadband} excitation signal so that maximum information about the behaviour in the band of excitation can be efficiently extracted. In this paper, to characterize the battery's short term electrical response at the different levels of SoC, while keeping the temperature constant at $25$\textdegree Celcius, we make use of a nonparametric characterization technique proposed by \cite{Relan2016}. It utilises the properties of the specially designed random phase multisine signal. Multisine excitation offer various advantages over the random Gaussian noise signals in extracting information from dynamical systems \cite{Bayard:1993, rivera2009constrained}. Detailed information on the use of multisine signals for identification can be found in \cite{RikJohanBook2012}.    

\begin{definition} Consider a signal $u$ with a power spectrum $S_U(j\omega)$,  which is piecewise continuous, with a finite number of discontinuities. A random signal belongs to the Riemann equivalence class of $u$ if it obeys by any of the following statements:
\begin{inparaenum}
 \item  It is a Gaussian noise excitation with power spectrum $S_U(j\omega)$.
 \item It is a random multisine or random phase multisine \cite{RikJohanBook2012} such that:
 \begin{equation}
 \frac{1}{N}\sum\limits_{k = k_1}^{k_2} \mathbb{E} \{|U(j\omega_{k})|^2\}= \frac{1}{2\pi}\int\limits_{\omega_{k_1}}^{\omega_{k_2}}S_U(\nu)d\nu + \mathcal{O}(N^{-1})
 \end{equation}
\end{inparaenum}
\end{definition}  where $\omega_{k} = k\frac{2\pi f_{s}}{N}, k \in \mathbb{N}, 0 <\omega_{k_{1}}<\omega_{k_{2}}< \pi f_s $ \textit{and} $f_s$ is the sample frequency. The  frequency domain representation of the multisine signal is the sum of the Fourier transforms of the individual sines and is given by:
\begin{equation}
U_{ms}(j\omega) = \frac{1}{\pi \sqrt{N_k}}\sum\limits_{k \in \pm \mathbb{K}_{exc}}{A}(k) \delta(\omega_{k}- \omega_{k_e}) e^{j \varphi_k}
 \end{equation} where $\delta(\bullet)$ is the Dirac delta function, $\mathbb{K}_{exc}$ is the index of excited frequencies, $N_k$ the number of excited frequencies and $\varphi_k$ are the phases. 
 The amplitudes of the multisine components ${A}(k)$ can be chosen arbitrarily, depending on the application. In the next section, we will briefly explain how the properties of a multisine excitation can be exploited to gain better insight into the dynamics of the system.

\section{\textbf{Nonparametric Characterization}}
\label{NonChar}
\begin{figure}[!ht]
\centering
\captionsetup{justification=centering}
\includegraphics[width=0.3\textwidth]{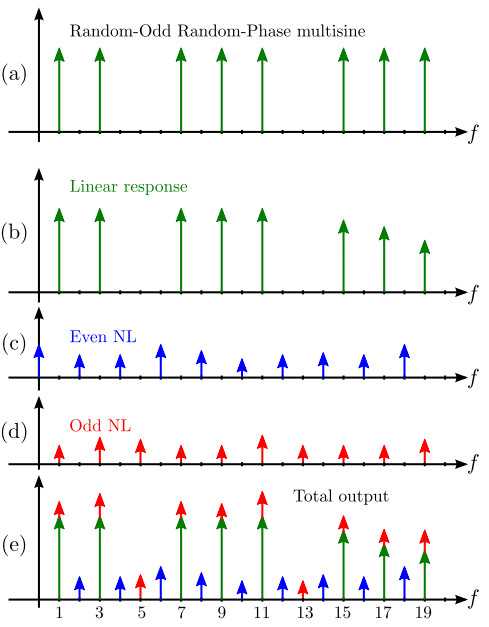}
\caption{Response PISPO: Total output is the sum of linear contributions (at excited lines), even NL (at even lines), odd NL (at odd lines) and noise (at all lines, not displayed here).}
\label{MultiNonResp1}
\end{figure}
\begin{definition}
For this particular analysis, it is assumed that the battery discharge capacity and the corresponding SoC levels are exactly known, or can be estimated accurately.
\end{definition}

\begin{definition} It is assumed that battery can be modelled as a weakly nonlinear periodic-in-same-period-out (PISPO) systems described by using Volterra series (see \cite{MartinSchetzen1980, BoydChua84, RikJohanBook2012} for more details). 
\end{definition}
\begin{remark} A nonlinear system is called PISPO if the steady state response to a periodic input is also a periodic signal with the same period as the input (with preservation of the period length). This includes systems with saturation and discontinuous nonlinearities, but it excludes systems with period multiplication, chaotic behavior, sub-harmonics, and hysteresis, see \cite{BoydChua84, BoydChua85, Schoukens1998param, Schoukens2009Nonparam} for a more formal definition.
\end{remark}
 In this section, a random odd multisine excitation signal \cite{Yves2009} with  a specially chosen $A(k_{n.exc}) = 0$ is used, see Fig.\ref{MultiNonResp1}(a). A linear system would only generate energy at the excited frequency lines, see Fig.\ref{MultiNonResp1}(b) and it would only consist of the green contributions. Whereas, a nonlinear system can also generate energy at non-excited lines: hereinafter termed as the the \textit{detection lines}. For example, a nonlinear system with degree $m$ can generate energy at any output frequency that is the sum of $m$ frequencies $f_i$ that are present in the input (spectrum), where repeated selection of the same frequency is allowed \cite{RikJohanBook2012}. Even nonlinearities only generate energy at even \textit{detection lines}, assuming that no constant term is present in the multisine. An even combination of odd lines is always even. As such, the level of the even nonlinearities can be quantified immediately by looking on the even lines in the output spectrum Fig.\ref{MultiNonResp1}(c). Similarly, the non-excited odd lines serve as \textit{detection lines} for the odd nonlinearities. This is visualised in Fig.\ref{MultiNonResp1}(d). Since the nonlinear system is operating in open-loop, the output Discrete Fourier Transform (DFT) spectrum of each period of the steady state response (with known input) to an odd random phase multisine with random harmonic grid is given by:
\begin{equation}
 Y^{[p]}(k) = Y_0(k)+ {N^{[p]}_Y}(k)+ Y_S(k)
\end{equation} The total response of the system is the sum of linear ($Y_0(k)$) and stochastic nonlinear (even \& odd) contributions $Y_S(k)$, where $p=1,2,3,...,P$, periods of the multisine and $N^{[p]}_Y$ is the noise term. This is depicted in the Fig.\ref{MultiNonResp1}(e) in the case of noiseless measurements. For interested readers. a detailed description of the procedure can be found in \cite{ RikJohanBook2012, Relan2016}. 

\subsection{\textbf{Observations: Nonparametric characterization}}
For the completeness of this section, we show here the observations made from the nonparameteric characterization of the Li-ion Polymer Battery (EIG-ePLB-C$020$, Li(NiCoMn)). For detailed information on the measurement set-up and experiment design, please see below section \ref{MeasExp}.
\begin{figure}[!ht]
 \centering
\captionsetup{justification=centering}
 \includegraphics[width=0.45\textwidth]{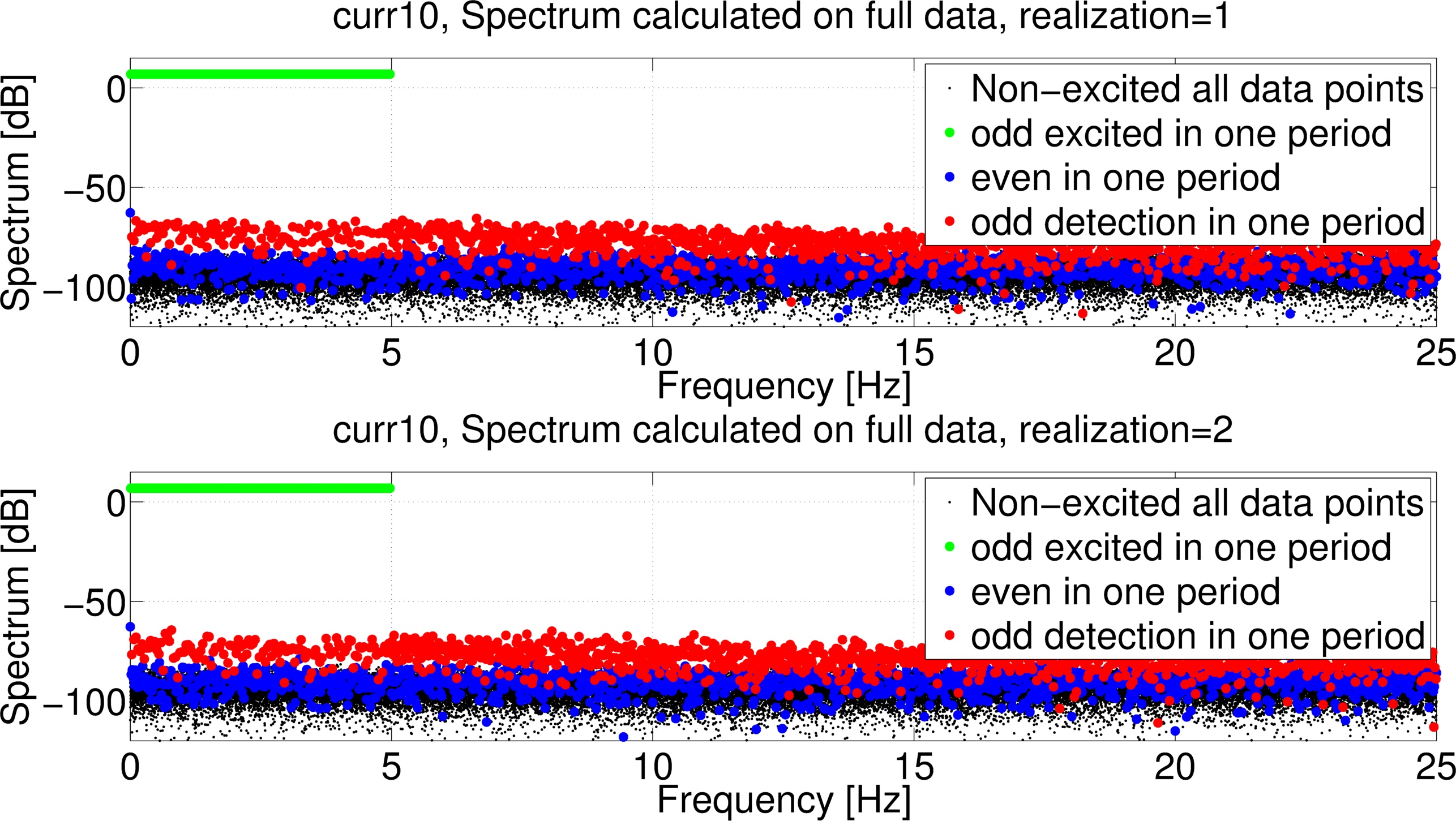}
\caption{Current profile in frequency domain at all SoC's}
\label{ExperimentCurrent}
\end{figure}

\begin{figure}[!ht]
 \centering
\captionsetup{justification=centering}
 \includegraphics[width=0.45\textwidth]{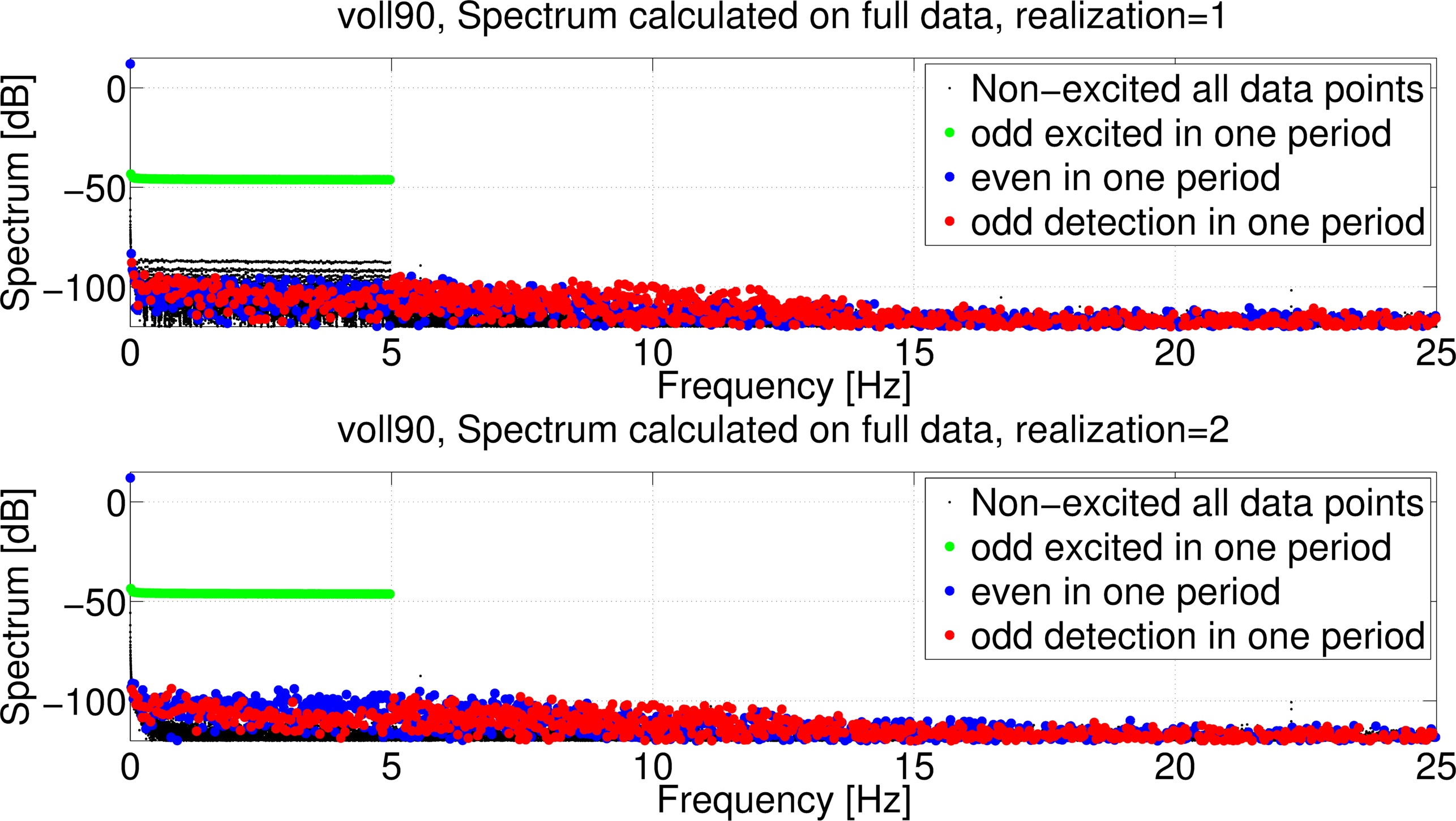}
\caption{Voltage response in frequency domain at 90\%  SoC}
\label{ExperimentVoll90}
\end{figure}

\begin{figure}[!ht]
 \centering
\captionsetup{justification=centering}
 \includegraphics[width=0.45\textwidth]{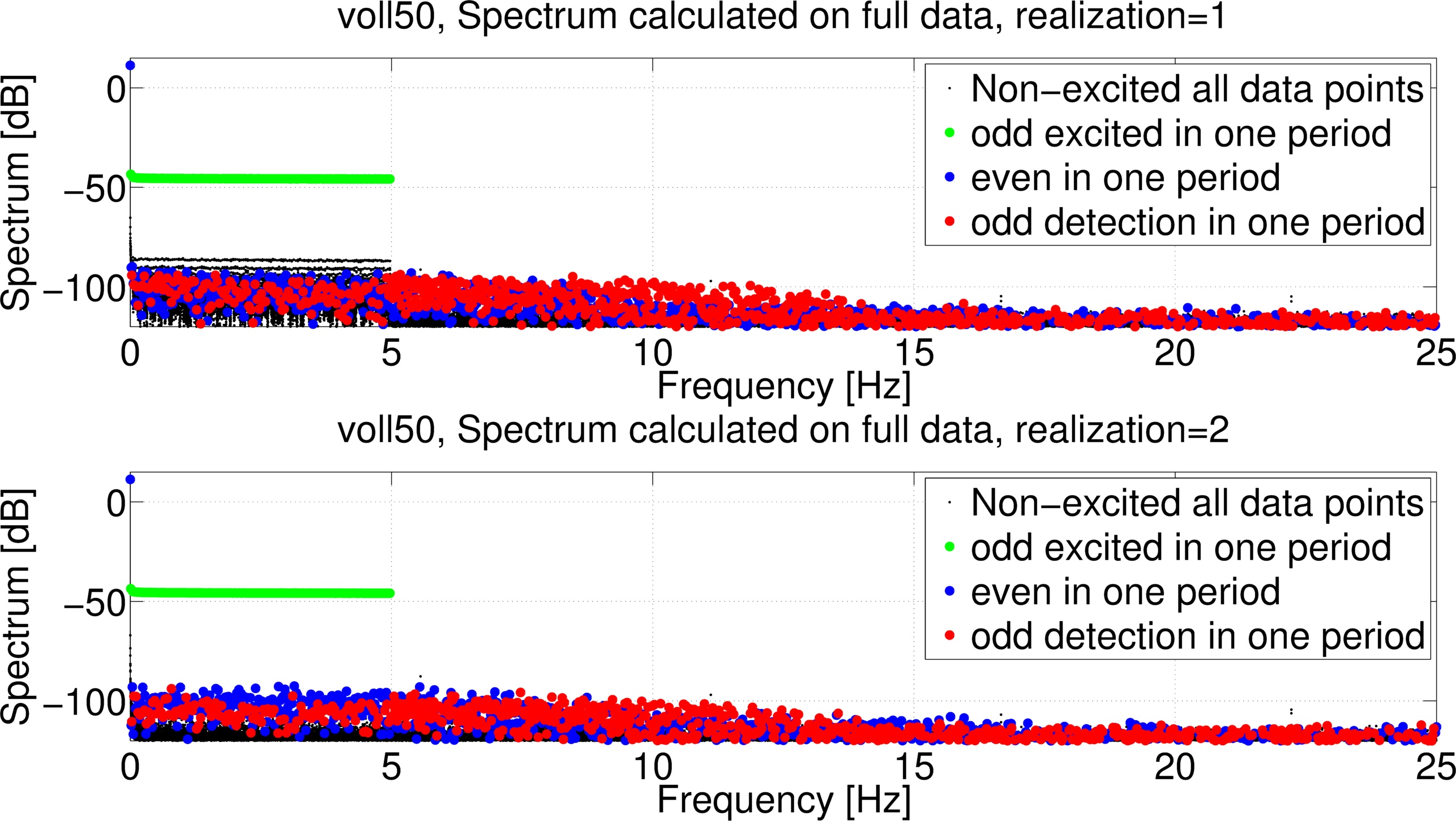}
\caption{Voltage response in frequency domain at 50\% SoC}
\label{ExperimentVoll50}
\end{figure}  
\begin{figure}[!ht]
 \centering
\captionsetup{justification=centering}
 \includegraphics[width=0.45\textwidth]{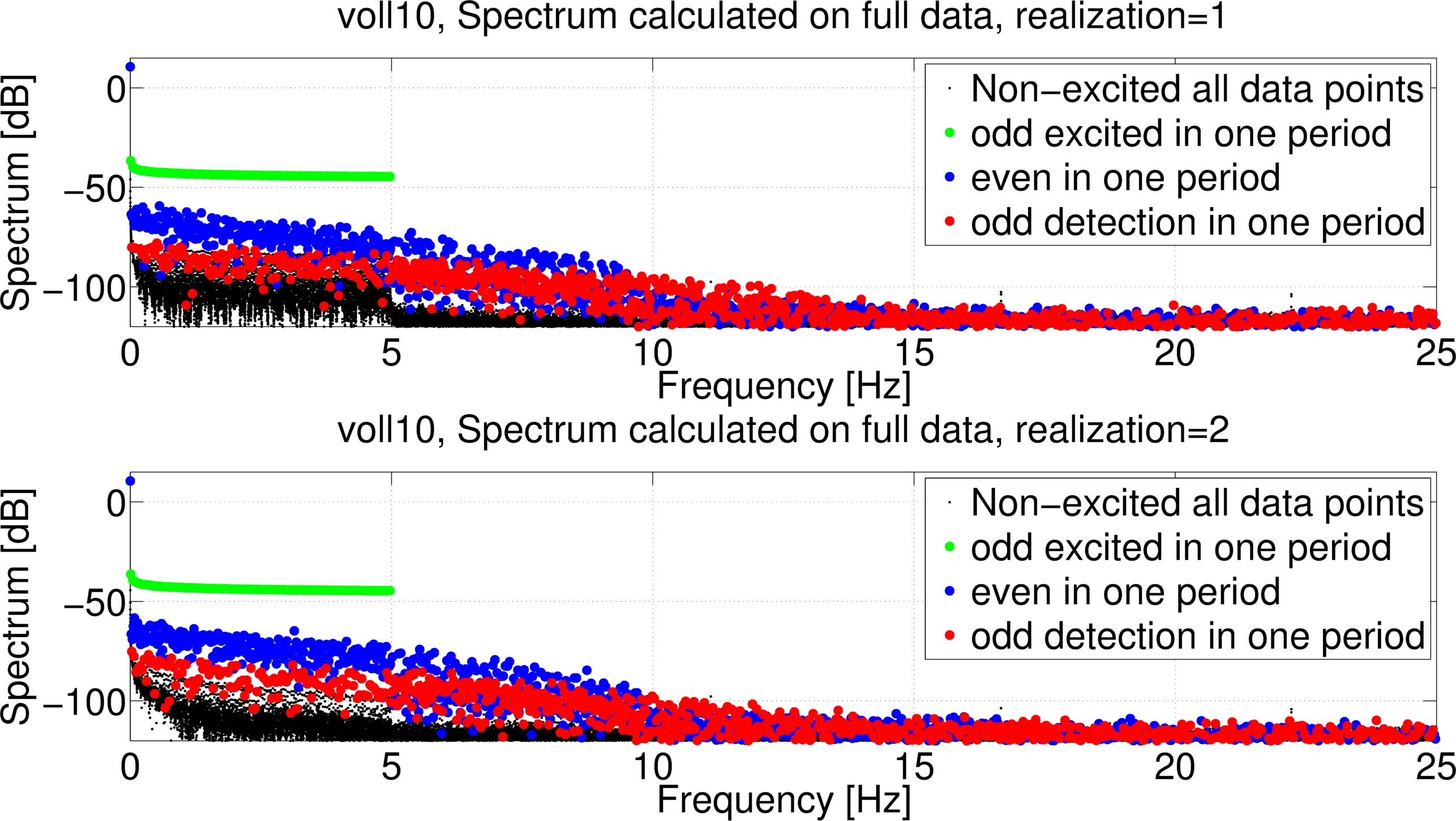}
\caption{Voltage response in frequency domain at 10\% SoC}
\label{ExperimentVoll10}
\end{figure} Both measured input and output signals are periodic in nature with the preservation of period length, hence the  PISPO assumption is validated. The odd-random phase multisine excitation current profile signal applied to the battery at all levels of SoC's is shown in the Fig.\ref{ExperimentCurrent}. It can be seen that the power is only injected in the frequency band of interest, which in this particular case lies between $1$Hz\textendash$5$Hz. A very small odd (red dots) nonlinear effect is visible at $-70$ dB. A probable cause may be the non-ideal behaviour of the the power electronics (e.g. switching of IGBTs) in the data acquisition system. In the odd random multisine current input load profile, only odd frequency lines are excited. In addition to that, in the frequency band of interest some of the randomly chosen odd frequency lines are not excited. This is done, in order to quantify also the contribution of odd nonlinear distortions along with the even nonlinear distortions. Hence, the term \textit{ "detection lines"} is used for the non-excited odd and even frequency lines. Using the methodology discussed above, valuable information about the nonlinear behaviour of the battery over its complete operating range w.r.t SoC levels at $25$\textdegree Celcius is extracted. Fig.\ref{ExperimentVoll90}, Fig.\ref{ExperimentVoll50}, and Fig.\ref{ExperimentVoll10} show the output voltage response of the battery at 90\% SoC, 50\% SoC and at 10\% SoC to the applied multisine excitation current signal for two different realizations respectively. It can be observed that the battery behaves almost linearly (green dots) at the 90\% SoC, and at the 50\% SoC. A similar behaviour was also observed at the 30\% SoC and 70\% SoC levels. At the 10\% SoC even (blue dots) and odd (red dots) nonlinear effects are visible, but dominantly even behaviour can be observed. The results obtained are in accordance with the behaviour of the \textit{Open circuit voltage (OCV)\textendash SoC} curve of the battery \cite{EIG-NMC20Ah}. According to the shape of the \textit{OCV\textendash SoC} curve between $10\%$ SoC\textendash$90$\% SoC, the battery behaviour is almost linear in the neighbourhood of the operating point at all temperatures, whereas at the $10$\% SoC level, the battery operating point is at the cusp of linear and nonlinear regime of its range of operation. As mentioned before, at 10\% SoC the nonlinear distortion become significant therefore a nonlinear model is necessary to capture efficiently the battery dynamics. 

The need for differentiating between odd and even frequencies during the nonparametric test is essential in order to get an idea about the contributions of nonlinear distortions to the FRF and to get an initial idea about the polynomial degrees in the PNLSS model which may be required to capture the nonlinear behaviour of the system. It can be seen from the battery nonparametric analysis that the contributions of both odd and even nonlinear distortions becomes significant at $10\%$ SoC, hence its gives us an early indication that both even and odd degree monomials will be required to capture the nonlinear effects. In the next section, a data driven nonlinear model identification technique is proposed once this information about the SoC dependent nonlinear operating regime of the battery is extracted using the nonparametric characterization.
\section{\textbf{Polynomial Nonlinear State-Space Models}}
\label{NonMod}
For most control as well as prediction problems a flexible, compact yet an easy-to-initialize model with an ability to describe Multiple-Input Multiple-Output (MIMO) systems is required. A state-space representation  of  the  system is often a good choice.  A  general $n_{a}^{th}$ order  discrete-time   state-space   model   is   described   by the following equations:
\begin{align}
 x(t+1) &= f(x(t),u(t))\nonumber\\ 
 y(t) &= g(x(t),u(t))
\label{eqn:NSS}
\end{align}with $u(t)\in \mathbb{R}^{n_u}$  the vector containing the ${n_u}$ inputs at time $t$, and $y(t)\in \mathbb{R}^{n_y}$  the vector containing the ${n_y}$ outputs. The state vector $x(t)\in \mathbb{R}^{n_a}$  represents the memory of the dynamical system. One of the ways in which a nonlinear state-space moel can be identified directly from the data is proposed in \cite{schon2011system}.
A nonlinear state-space model where $f(\cdot), g(\cdot)$ are approximated by polynomial basis functions is an another representation for a MIMO system. A PNLSS model structure \cite{Paduart2010} is very flexible to capture both nonlinear feed-forward and feedback (e.g. shifting resonances) dynamics. It is very easy to initialize it via Best  Linear  Approximation (BLA). The PNLSS model can be described as:
\begin{align}
 x(t+1) &= Ax(t) + Bu(t) + E\zeta(t)\nonumber \\
 y(t) &= Cx(t) + Du(t) + F\eta(t)+e(t)
\label{eqn:PLNSS}
\end{align}The coefficients of the linear terms in $x(t)\in \mathbb{R}^{n_a}$   and $u(t)\in \mathbb{R}^{n_u}$  are given by the matrices $A \in \mathbb{R}^{n_a \times n_a}$ and $B \in \mathbb{R}^{n_a \times n_u}$ in the state equation,  $C \in \mathbb{R}^{n_y \times n_a}$ and $D \in \mathbb{R}^{n_y \times n_u}$ in the output equation. The  vectors $\zeta(t) \in \mathbb{R}^{n_{\zeta}}$ and $\eta(t) \in \mathbb{R}^{n_{\eta}}$ contain nonlinear monomials in $x(t)$ and $u(t)$ of  degree  two  up  to  a  chosen  degree  $P$ .  The coefficients of nonlinear terms are given by the matrices $E \in \mathbb{R}^{n_a \times n_{\zeta}}$ and $F \in \mathbb{R}^{n_y \times n_{\eta}}$.  

\section{\textbf{Identification procedure of the PNLSS}}
\label{IdenPNLSS}
The structure of the black-box state-space model given in (\ref{eqn:PLNSS}) lends itself to an efficient, three major steps identification procedure described below: 
\begin{enumerate}[A)]
\item First, initial estimates of the $A, B, C$ and $D$ matrices are obtained. In order to do so, a nonparametric estimate of the system's frequency response function (FRF) is determined in mean square sense. This is called the BLA.
\item Then, a parametric linear model (linear subspace $A,B, C, D$ matrices) is estimated from this nonparametric BLA. Thereafter, the subspace estimates are optimised in maximum likelihood sense by applying a nonlinear minimisation routine.
\item Finally the full nonlinear model, including the polynomial coefficients is estimated.                                                                                                                                                                                                                                              
\end{enumerate}
The complete procedure is carried out in the frequency domain, which opens the possibility to apply user-defined weighting functions in specific frequency bands. 
\subsection{\textbf{Best Linear Approximation}}
\begin{Def}    
The Best Linear Approximation (BLA) of a nonlinear system is defined as the model $G$ belonging to the set of linear models $\mathcal{G}$, such that
\begin{equation}
 G_{BLA}= \underset{G \in \mathcal{G}}{\operatorname{arg\,min}} \hspace{0.1cm} \mathbb{E} \left( |y(t)-Gu(t)|^2 \right)
\end{equation}
\end{Def} 
\paragraph*{Set Up}
\begin{figure}[!h]
\centering
\captionsetup{justification=centering}
\includegraphics[width=0.4\textwidth]{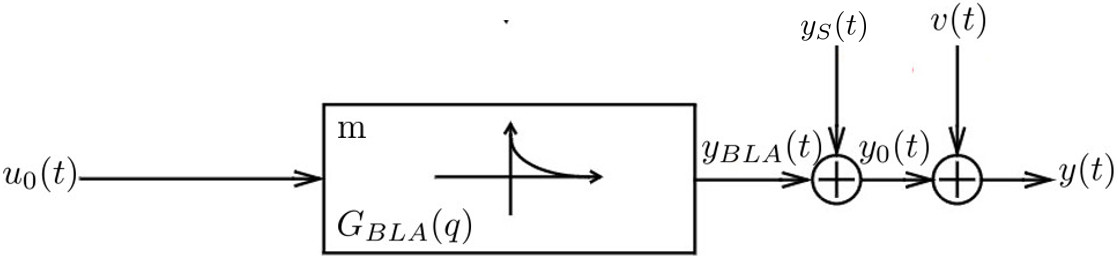}
\caption{ Time domain representation of the problem}
\label{MultiNonResp}
\end{figure}Here, we focus for simplicity on the estimation of the Best Linear Approximation (BLA) $G_{BLA}(q)$ of a discrete time single-input-single-output (SISO) model of a nonlinear system $\textbf{\textit{m}}$  nonparametrically, which is excited with signals belonging to the Riemann equivalence class of asymptotically normally distributed excitation signals \cite{RikJohanBook2012}, see Fig. \ref{MultiNonResp}. For an infinitely long data record $t = -\infty , . .. , N -1,$ the input-output relation of the nonlinear system can be written as:
\begin{equation}
y (t) = G_{BLA}(q)u_0(t) + y_s(t)+ H_0(q)e(t). 
\end{equation} 
with $q^{-1}$ the backward shift operator $(q^{-1}x(t) = x(t-1))$ and $y_s(t)$ are the stochastic nonlinear contributions. The exact input $u_0(t)$ is assumed to be known, while the output is disturbed with additive noise $v(t)$, then $y (t) = y_0(t) + v(t)$. The noise $v(t)$ is assumed to be filtered white noise, $v(t) = H_0(q)e(t)$, where $H_0(q)$ is the noise model. For a finite record length $t = 0, . . . , N -1$, as it is in practical applications, this equation has to be extended with the initial conditions, or in other words, the  transient effects of the dynamic plant and noise system $t_G$ , $t_H$:
\begin{equation}
y(t) = G_{BLA}(q)u_0(t) + y_s(t)+ H_0(q)e(t)+ t_G(t) + t_H(t). 
\label{TransTermTime}
\end{equation} Using the discrete Fourier transform (DFT)
\begin{equation}
 X(k)= \frac{1}{\sqrt{N}}\sum\limits_{t = 0}^{N-1}x(t) e^{-j2{\pi}kt/N},
\end{equation} an exact frequency domain formulation of (2) is obtained:
\begin{align}
 Y(k) &= G_{BLA}(\omega_{k} ) U_0(k) + Y_s(k) + H_0(\omega_{k}) E(k)\nonumber \\
      &+ T_G (\omega_{k}) + T_H (\omega_{k}) 
 \label{fullLPM}
\end{align} where the index $k$ points to the frequency $kf_s/N$ , with $f_s$  the sampling frequency, and $\omega_{k}=  e^{j2{\pi}kf_s /N} $. The finite record length requires the use of transient terms in (\ref{TransTermTime}), and it turns out that the leakage errors of the DFT are modelled by very similar terms in the frequency domain \cite{RikJohanBook2012}.
It is most important for the rest of this paper to understand that (\ref{fullLPM}) is an exact relation, where leakage effects are modelled by the transient terms \cite{pintelon1997frequency,  RikJohanBook2012}.  All these terms $t_G(t) , t_H(t) , T_G(k) , T_H(k) $ are described by rational forms in $q^{-1}$ (time domain) or $z^{-1}$(frequency domain), hence they are smooth functions of the frequency. Within the above described set up, the nonparametric BLA can be calculated using either the Fast or the Robust method explained in \cite{RikJohanBook2012}, or the Local Polynomial Method (LPM), which makes an optimal use of the smooth behaviour of $G_{BLA}$ and $T_G$ to significantly reduce the leakage errors \cite{ Schoukens2009Nonparam}. This results in superior properties compared with the classical windowing methods and provides a good estimation of the BLA as well as its variance ($\sigma^2_{BLA}$) \cite{Schoukens2009Nonparam}. An alternative to the LPM method to capture transient effect is \textit{TRansient Impulse response Modeling Method (TRIMM)} proposed by \cite{gevers2012transient}.
\subsubsection{\textbf{Nonparametric Identification Using LPM method}}
\label{NonParam}
In this section, we give a very brief introduction to the LPM method, which is used to estimate nonparametrically the FRF from the input current and the output voltage data. A detailed description, together with a full analysis is also given in \cite{pintelon2010estimation1, pintelon2010estimation2}, a comparison with the classical spectral windowing methods is found in \cite{Schoukens2009Nonparam}. The basic idea of the LPM method is quite simple:  the transfer function $G_{BLA}$, and the transient term $T_G$ are smooth functions of the frequency. So they can be approximated by a complex polynomial in a narrow band of frequency around a user specified frequency $k$ . The complex polynomial parameters are directly estimated from the experimental data.  Next $G_{BLA}(k)$, at the central frequency $k$, is retrieved from this local polynomial model as the measurement of the FRF at that frequency. This step is repeated every time for all DFT frequencies in the band of interest by shifting the sliding window over one DFT bin. Hence a local estimate of the FRF is obtained at every frequency.
Consider the full output error-expression described by (\ref{TransTermTime}), and an equivalent relation for the DFT-spectra applied to both the plant $G_{BLA}(q)u_0(t)$ and the noise term $v(t)= H_0(q)e(t)$:
\begin{align}
 Y(k) &= G_{BLA}(\omega_{k} ) U_0(k) + Y_s(k) + H_0(\omega_{k}) E(k)\nonumber \\
       &+ T_G (\omega_{k}) + T_H (\omega_{k})\nonumber \\  
 &= G_{BLA}(\omega_{k})U_0(k) + T (\omega_{k}) + V_0(k) + Y_s(k)
 \label{eqn:OEE}
\end{align} where the generalized transient term $T(\omega_{k}) = T_G(\omega_{k} ) + T_H(\omega_{k} )$
accounts for the leakage of the plant and noise dynamics.  The remaining noise term is $V_0(k) = H_0(\omega_{k})E(k)$. Making use of the smoothness of $G _{BLA}$ and $T$ , the following Taylor series representation holds for the frequency lines $k + r$, with $r = 0, \pm 1, . . . , \pm n.$ 
\begin{align}
 G_{BLA}(\omega_{k+r}) &=  G_{BLA}(\omega_{k})  + \sum\limits_{s = 1}^{R} g_s(k)r^s + \mathcal{O} \left(\frac{r}{n} \right)^{{R+1}} \\
 T(\omega_{k+r}) &= T(\omega_{k}) +  \sum\limits_{s = 1}^{R} t_s(k)r^s + N^{\frac{-1}{2}} \mathcal{O} \left(\frac{r}{n} \right)^{{R+1}}
\end{align} Putting all parameters $G_{BLA}(\omega_{k} )$, $T (\omega_{k})$ and the parameters of the Taylor series $g_s,t_s, s = 1, . . . , R$ ,  in a column vector $\theta$, and their respective coefficients in a row vector $K(k, r) $ allows (\ref{eqn:OEE}) to be rewritten (neglecting the remainders) as:
\begin{equation}
 Y(k + r) = K(k, r) \theta + V_0(k),
 \label{eqn:LPM}
\end{equation}Collecting (\ref{eqn:LPM}) for $r = -n, -n + 1, . . . , 0, . . . , n $ finally gives
\begin{equation}
 Y_n = K_n \theta + V_n ,
\end{equation} with $Y_n, V_n, K_n$ the values of $Y(k + r), V_0(k + r), K(k, r),$ stacked on top of each other.  Observe that the matrix $K$ depends upon $U_0$. Solving this equation in least squares sense eventually provides the best polynomial least squares estimate  $\hat{G}_{poly_{BLA}}(\omega_{k})$ for $G_{BLA}(\omega_{k})$. To get a full rank matrix $K_n$, enough spectral lines should be combined: $n \geq R + 1$. Least interpolation error is obtained for $n = R + 1 $ \cite{pintelon2010estimation1, pintelon2010estimation2}.%
\subsection{\textbf{Parametric BLA}}
Using the nonparametric FRF estimate ($\hat{G}_{poly_{BLA}}(\omega_{k})$) and its variance (${{\sigma}^2}_{\hat{G}{poly_{BLA}}}$) estimated in the last step, a parametric model $\hat{G}_{{BLA}_{par}}(q,\theta)$ is identified in least square sense to the $\hat{G}_{poly_{BLA}}$ using  the \textit{fdident} package \cite{fdident}. This discrete-time model describes the system as a rational transfer function. The model considered here is a rational function in the backward shift operator $q^{-1}$:
\begin{equation}
 \hat{G}_{{BLA}_{par}}(q,\theta) = \frac{b_0+b_1q^{-1}+b_2q^{-2}+......+b_{n_b}q^{-n_b}}{a_0+a_1q^{-1}+a_2q^{-2}+......+a_{n_a}q^{-n_a}},
\label{eqn:paraEq}
\end{equation} The parameter vector $\theta \in \mathbb{R}^{(n_b+n_a+2)*1}$  contains the parameters. Since one parameter can be chosen freely because of the scaling invariance of the transfer function, only $n_b + n_a + 1$ independent parameters need to be estimated. The order of the parametric model in (\ref{eqn:paraEq}) can be determined using the minimum description length (MDL) criterion \cite{RikJohanBook2012}. A balanced state-space realization $(G_{ss} = A,B,C,D)$ is calculated from the stable portion of the linear system $G_{{BLA}_{par}}(q,\theta)$, here the subscript $ss$ stands for the state-space and $A,B,C,D$ stand for the $A,B,C,D$ matrices in (\ref{eqn:PLNSS}). For stable systems, this is an equivalent realization for which the controllability and observability Gramians are equal and diagonal \cite{Laub1987}.
\subsubsection{\textbf{Nonlinear optimization of the BLA}}
Although the process described in the previous section uses numerically stable and efficient algorithms to retrieve the $A,B,C,D$ matrices, but in the case of output measurements with a low signal-to-noise ratio, consistency of the initial state-space matrices $A, B, C, D$ is lost. Unbiased parameter estimates can be recovered by optimising the subspace model in maximum likelihood (ML) sense. The ML framework also guarantees asymptotically  the lowest possible uncertainty on the model parameters, i.e. the efficiency of the estimates \cite{RikJohanBook2012}. A ML estimate can be obtained by minimizing the following cost function:
\begin{equation}
 V_{ss}(\theta) = \sum\limits_{k = 1}^{F} \frac{|\hat{G}_{BLA}(j\omega_{k} ) - G_{SS}( A, B, C, D, z_k )|^2}{{\sigma_{BLA}^2}(j\omega_{k})}
\end{equation} with $\theta = [ vec^{T}( A ) ; vec^{T}( B ) ; vec^{T}( C ) ; vec^{T}( D ) ]^T$ and $z_k = e^{j\frac{2\pi k}{N}}$ and $N$ is number of points per period. ${\sigma_{BLA}^2}(j\omega_{k})$ includes both noise and nonlinear distortion. Therefore, the parameters $\theta$ are then used as starting values in a nonlinear optimization of $V_{ss}$ with respect to (w.r.t) $\theta$, where the subscript $ss$ stands for the state-space. 
\subsection{\textbf{Estimation of the full nonlinear model}}
\subsubsection{\textbf{Trend removal}} The battery can be considered as a dynamic system with an integrating effect \cite{widanage2015estimating}. In addition to that a systematic shift in the data statistical property can result from sensor drift, non-ideal behaviour of data acquisition system etc. Signal drift is considered a low-frequency disturbance and can result in unstable models. Therefore to remove the nonstationary effects from the data as well as to improve the model performance at the low frequencies, before the estimation of the complete nonlinear model, the underlying trend is removed using the $\ell_{1}-$regularized trend removal technique developed in \cite{kim2009ell_1}. The trend estimate $m$ as the minimizer  of the weighted sum objective function can be defined as:
\begin{equation}
 \frac{1}{2}\norm{{y-m}}^2_{2}+\lambda \norm{Dm}_1
\end{equation} where the trend estimate $ m = (m_1,m_2,.........m_n) \in \mathbb{R}^{n}$, the battery output $y = (y_1,y_2,.........y_n) \in \mathbb{R}^{n}, \norm{c}_i = \sum_{i}|c_i|$ denotes the $\ell_{1}-$ norm of the vector $c$ and $D \in \mathbb{R}^{(n-2) \times n}$ is the second-order difference matrix (which is Toeplitz in nature)\cite{kim2009ell_1}. The first term in the objective function measures the size of the residual whereas the second term measures the smoothness of the estimated trend. The $\ell_{1}-$trend method produces trend estimates that are piecewise linear, and therefore it is well suited to analyze battery time series data, which can be thought of having a slowly time-varying system with underlying piecewise linear trend. 
\subsubsection{\textbf{Estimation of full PNLSS}} In the last step, the coefficients of both the linear and the nonlinear terms in (\ref{eqn:PLNSS}) are identified. 

\begin{definition}    
It is assumed that the input $u(t)$ of the model in Section \ref{NonMod} is noiseless, i.e., it is observed without any errors and independent of the output noise.
\end{definition} 

\begin{definition}    
The nonlinearity is assumed to be smooth and can be approximated well using polynomial basis functions.
\end{definition}

\begin{remark} A uniformly convergent polynomial approximation of a continuous nonlinearity is always possible on a closed interval  due  to  the  Weierstrass  approximation  theorem. The type of convergence can be relaxed to mean-square convergence to allow some discontinuous nonlinearities as well.
\end{remark}
For the identification of full PNLSS, a weighted least squares approach is employed to keep the estimates of the model parameters unbiased. The Weighted Least Squares (WLS) cost function that needs to be minimized with respect to the parameter $\theta_{NL} = [vec^{T}(A); vec^{T}(B); vec^{T}(C); vec^{T}(D); vec^{T}(E);$ $vec^{T}(F)]^T$  is given by:
\begin{equation}
 V_{WLS}(\theta_{NL}) = \sum\limits_{k = 1}^{N_t} \frac{|Y_{mod}(j\omega_{k},\theta) - Y(j\omega_{k})|^2}{W(j\omega_k)}
 \label{eqn:NonEst}
\end{equation} where $N_t$ is the total number of selected frequencies. $Y_{mod}$ and $Y$ are the DFTs of the modelled output and the measured output, respectively. Because in nonlinear systems, model errors often dominate the disturbing noise, we put the weighting factor $W(j\omega_k) = 1$. Only if the model errors are below the noise level, $W(j\omega_k) $ can be put equal to the noise variance ${\sigma_{n}}^2(j\omega_{k})$. Furthermore, the model error $\epsilon (j\omega_{k},\theta_{NL}) \in \mathbb{C}^{n_y}$  is defined as
$\epsilon (j\omega_{k},\theta_{NL}) = Y_{mod}(j\omega_{k},\theta_{NL}) - Y(j\omega_{k})$, 
The minimization of the non-convex cost function (\ref{eqn:NonEst}) is tackled via the Levenberg-Marquardt scheme \cite{more1978levenberg}. To ensure good initial values, the $G_{ss}$ is used for initialize the nonlinear model. Hence, the identified full PNLSS model cannot perform worse than the BLA in least squares sense. In addition, the performance of the PNLSS model also depends on the choice of degree of nonlinearity $P$. A proper choice of degree $P$ is a trade-off between the model complexity and its performance on the validation data. Here, a polynomial degree up-to $3$ has been selected for both state and output equations. Other ways of initializing the PNLSS model are proposed in \cite{van2013comparison, marconato2014improved}. 

\section{\textbf{Measurement Setup and Experiment Design}}
\label{MeasExp}
A high energy density Li-ion Polymer Battery (EIG-ePLB-C$020$, Li(NiCoMn)) with the following electrical characteristics: nominal voltage $3.65 V$, nominal capacity $20$Ah, AC impedance ($1$kHz) $< 3m\Omega$ along with the PEC battery tester SBT0550 with $24$ channels is used for the data acquisition. The tests are performed on a pre-conditioned battery inside a temperature controlled chamber at $25$\textdegree Celcius. An odd-random phase multisine signal is used as an input excitation signal. The band of excitation is kept between $1$Hz\textendash$5$Hz, because the dynamic range of interest of the battery for HEV's and EV's applications is covered well within this band of excitation. It also takes into consideration the limitations of the battery tester in terms of the sampling frequency.   The excitation signal has a period of 5000 samples and the sample frequency $f_s$ is set to $50$Hz resulting in a frequency resolution of $f_o = 0.01$Hz. The range of excitation frequency is also limited due to the system limitations of the PEC testers. The input is zero mean with a rms value of $20$A. Two different realizations with random realization of the phases of the multisine signal with $20$ periods are acquired at different levels of SoC's. For the test, the battery is first charged using a constant $\frac{C}{3}$ rate, where $C$ is the rated capacity, to the maximum charge voltage of $4.1$V using the constant current-constant voltage method. Then, after a relaxation period of $30$ minutes, it is discharged to the desired SoC level Ah-based and considering the actual discharge capacity at $25$\textdegree Celcius until the end of discharge voltage $3.0$V of the cell. After each discharge a rest period of $60$ minutes is applied before the multisine tests are performed. It is made sure that the synchronisation is maintained between the signal generation and acquisition side.  
\section{\textbf{Results}}
\label{res}
\subsection{\textbf{Nonlinear modelling}}
\begin{figure}[!ht]
 \centering
\captionsetup{justification=centering}
 \includegraphics[width=0.4\textwidth]{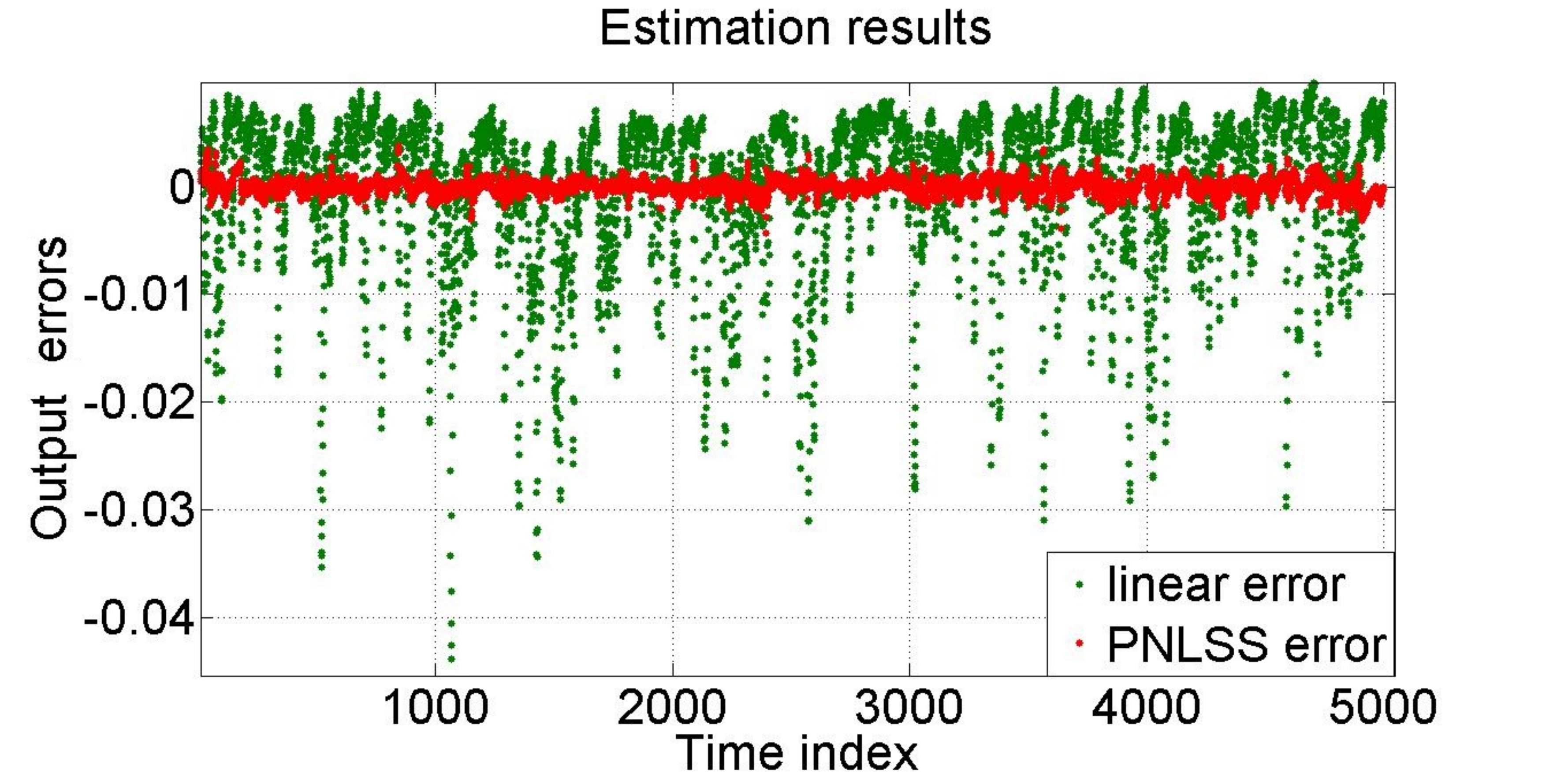}
\caption{Model responses (zoomed Time domain) at $10$\%SoC}
\label{ExpNonLinTimeSoc10Zoom}
\end{figure}

Although the PLNSS structure is capable of capturing influence of the SoC, the current level and the temperature in its MIMO settings, we validate in this study its usability at one particular operating condition of $25$\textdegree Celcius, $20$A rms current input, and $10\%$ SoC. 
\begin{figure}[!ht]
 \centering
\captionsetup{justification=centering}
 \includegraphics[width=0.4\textwidth]{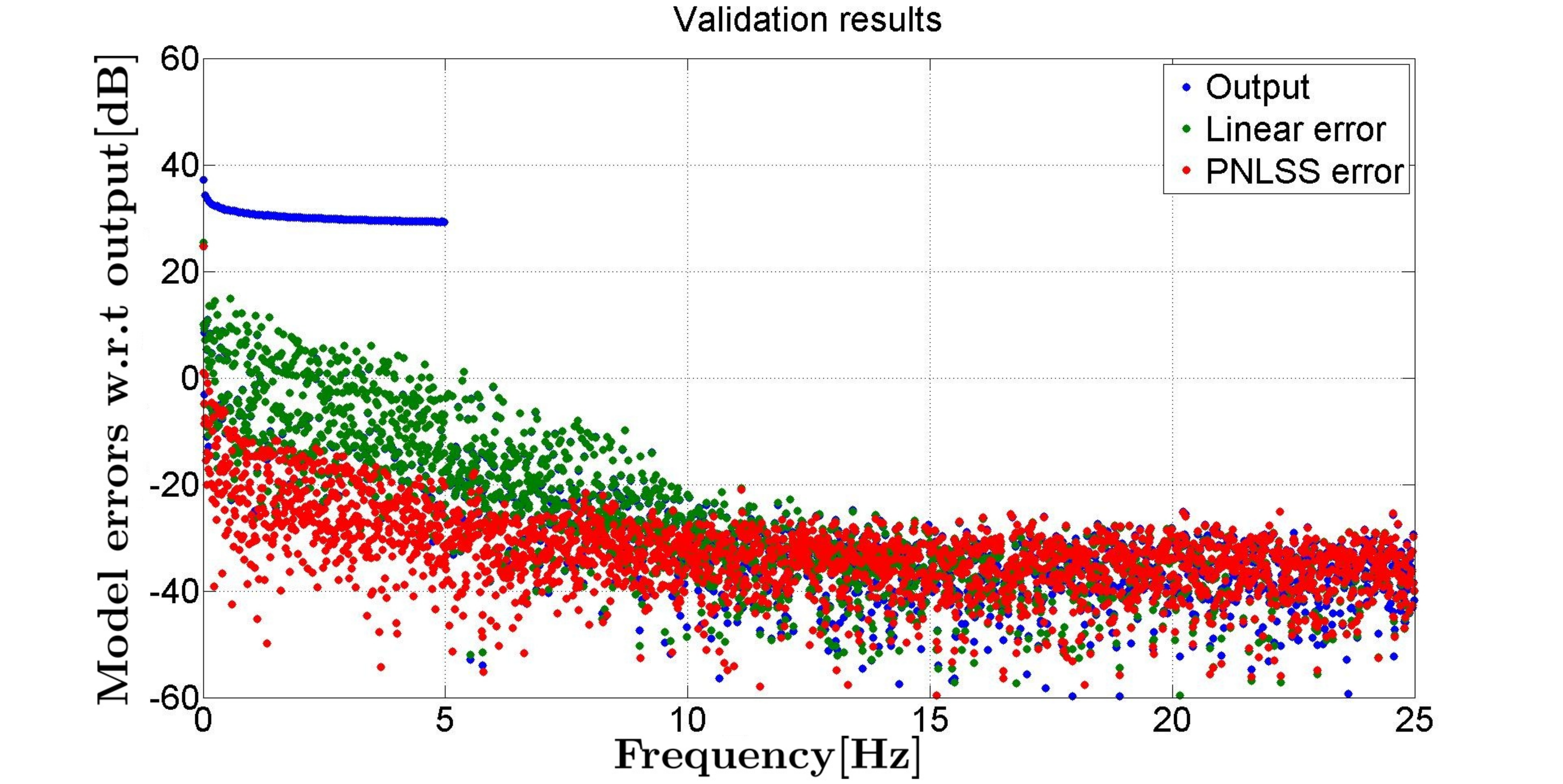} 
\caption{Errors at $10$\%SoC w.r.t the output (frequency domain)}
\label{ExperimentNonLinSoc10}
\end{figure}                
\begin{figure}[!ht]
 \centering
\captionsetup{justification=centering}
 \includegraphics[width=0.4\textwidth]{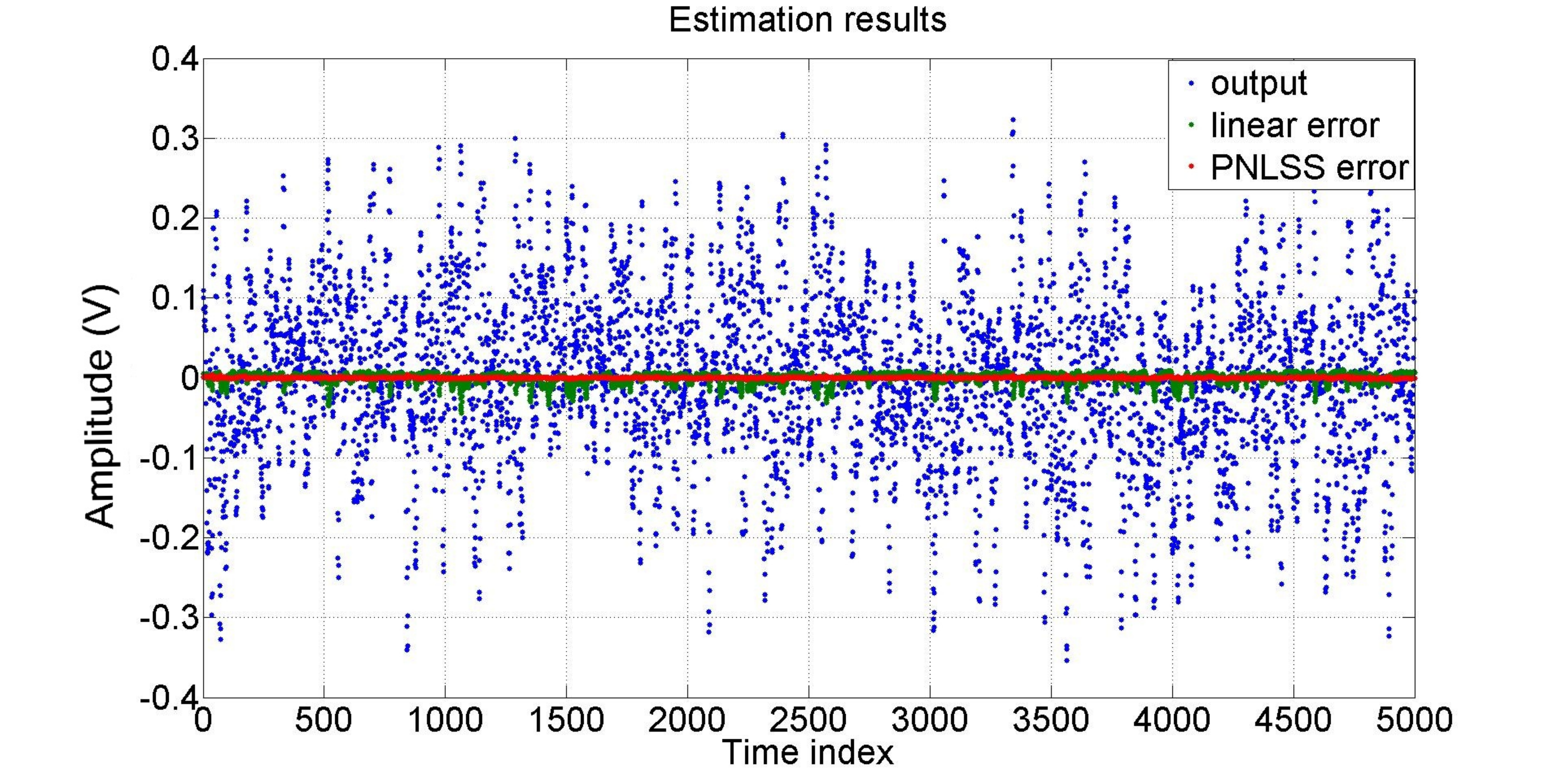}
\caption{Errors w.r.t the output (time domain) at $10$\%SoC, RMSE PNLSS $ = 2.8591\times 10^{-4}$, RMSE Linear $= 0.0072$}
\label{ExperimentNonLinTimeSoc10}
\end{figure}
Fig.\ref{ExperimentNonLinSoc10} and Fig.\ref{ExperimentNonLinTimeSoc10} show the comparison between the output responses of linear model and the PNLSS model in the frequency and the time domain respectively. The advantage of using the PNLSS becomes more clear after zooming in to the time domain response as shown in the Fig.\ref{ExpNonLinTimeSoc10Zoom}.
It can be clearly observed both from the frequency and the time domain plots that the PNLSS model structure is powerful and flexible enough to capture the dynamics of the battery. It outperforms the linear model by a factor of $10$ ( $ \simeq 20$ dB difference in the frequency band of interest) on the output error side, which is quite a significant achievement considering the fact the battery operates on the cusp of its nonlinear regime.
\section{\textbf{Conclusion}}
\label{conc}
In this paper, we proposed a very powerful yet simple PNLSS model structure for modelling the battery's nonlinear dynamics based on the information gained from the nonparametric characterization of battery's electrical response. Nonparametric characterization allows the detection and quantification of the nonlinear effects from the input-output measurements. This generic approach can easily be applied at different scenarios, e.g. at different SoC/SoH levels, temperatures, levels of charging and discharging currents. The PNLSS model structure outperformed the linear model by a factor $10$ on the output error. This is quite a significant achievement as the future aim of the battery manufacturers and industries is to push battery operation much deeper into its operational regime. 
\bibliographystyle{IEEEtran}
\bibliography{IEEEtranBibOld}

\end{document}